\documentclass[12pt,letterpaper,onecolumn]{IEEEtran}
\usepackage{setspace}
 \usepackage{xargs}

\usepackage{graphicx}
 \usepackage{amsfonts}
 \usepackage{amsmath}
\usepackage{amsthm}
\usepackage{authblk}
\usepackage{epstopdf}
\usepackage{url}
\usepackage{amssymb}
\usepackage{bbold}
 \usepackage{wasysym}
\theoremstyle{plain}

\theoremstyle{definition}

\usepackage[margin=1.2in]{geometry}

\begin{document}

\title{New Generation Value Networks for Content Delivery}

%

\author[$\dag$]{Stefano Iellamo}
\author[*]{Guenter Klas}
\author[*]{Kevin Smith}

\affil[$\dag$]{
siellamo@ics.forth.gr}
\affil[*]{
first\_name.last\_name@vodafone.com}

\maketitle


\begin{abstract}
In this paper we paint a broad picture of the Internet content delivery market, by taking into consideration 
both economical and technical challenges that might drive the interactions among the stakeholders in the future.
We focus on a few disrupting factors, namely ubiquitous encryption, traffic boost, network scalability, latency needs 
and network control; and try 
to figure out whether the current model (CDN) is robust against their variation, which optimization can be envisioned and how the 
most accredited option (ICN) can be of help.
\end{abstract}


\section{Introduction}

The original end-to-end Internet architecture has not been able to solve the scalability and coordination problems related 
to achieving 
quality through caching in content delivery, which has led to the fragmentation of the Internet to overlay networks, such as
Content Delivery Networks (CDNs), which can be thought of as large distributed systems of servers deployed in multiple data 
centers across the Internet that make use of smart caching operations to minimize latency.\\

However, Internet traffic and content volumes keep growing and caching importance and CDNs market are increasing accordingly.
A concrete example of the impact of caching comes from Uganda and Kenya, where the introduction of
Google Global Cache increased users' Internet traffic volume by $300-1000\%$ in only 2 weeks\footnote{From this one can
infer that in Uganda and Kenya users were interested in viewing video, 
but backhaul 
failures had previously prevented most videos to be played.}.
As a side effect, the growth of the CDN market is now challenging the traditional transit and peering models, creating
the need for a more uniform and efficient content delivery model better integrated with the Internet architecture. Information-
centric networking (ICN), a new networking paradigm, utilizes in-network caching and adapts the network architecture to
the content-driven network usage patterns. The concept
introduces routing based on unique content names instead of content location. 
In content delivery, CDNs utilize
distributed caching closer to the end-user and focus on content instead of location, and thus can be considered as a
predecessor of ICN. \\


To complicate things,
Snowden's revelations have caused a shift of attention towards 
the protection of users' privacy against attacks and pervasive monitoring. This has caused in turn
an unprecedented boost of encrypted contents (in the form of HTTPS traffic) between Web browsers 
and content servers which is accomplished by means of HTTPS sessions, i.e., a secure TLS tunnel is created between origin 
and destination. If on the one hand the use of HTTPS addresses most privacy concerns, on the other hand it has raised other alerts:
\begin{itemize}
   \item ISPs (here including Internet Access Providers and Backbone Access Providers) are almost powerless facing encrypted traffic. 
Monitoring and optimization activities cannot be performed except by terminating the HTTPS connection for inspection. This can be however
considered as a workaround (it is performed by means of a man-in-the-middle approach) and, furthermore, it adds delays due to processing
time which is impractical for large volume of network traffic \cite{le_quoc_scalable_2013}. 
\item CDNs which are also Content Providers (CPs) have the power to prioritize their traffic over that of other CPs
and a Net Neutrality regulation in this respect is actually missing. 
 \item HTTPS overhead in terms of latency is still considerable. This is due to the SSL initial key exchange and
to the CDN's difficulty in optimizing
caching operations . 
 \end{itemize}

Hence, if on the one hand ISPs are tempted to engage in tighter collaborations with the CDNs, on the 
other hand they see the CDNs as the barbarians at the gate who are attempting to conquer the last corner of the Net
by means of a Trojan horse. 
Thus, it makes sense that ISPs try to keep a certain control 
on the QoS experienced 
by the customers they serve and interface with. To this end, control and caching operations over encrypted video 
contents are of crucial
importance since encrypted videos account for a considerable (increasing) portion of the total Internet traffic (with
the recent move of Netflix to HTTPS, encrypted videos will soon represent more than half of the total Internet traffic in 
the US \cite{_most_????}). \\

One could argue that HTTPS is often too radical and an authenticity guarantee would be enough for many types of 
contents (e.g., YouTube videos). From this perspective, ICN would provide data security 
by design instead of tunnel security, meaning that data integrity (data 
has not been tampered with) as well as authentication (data is 
from who it says it is from) are ensured by signing data at creation\footnote{Note that confidentiality is not ensured 'by design' in ICN, though
proposals have been made in the literature to enable content-based confidentiality.}.  \\

Nevertheless, ICN is often looked at as 
a disruptive solution which might compromise the business of CDNs. 
With this in mind, we build on \cite{zhang_value_2014} and study the incentives for different network players
to engage in ICN and discuss protocol design considerations and policy interventions that might be necessary to ensure 
a positive socio-economic outcome from deploying ICNs (section \ref{sec:icn}).\\

In the following we will focus on the interconnection layer\footnote{The Interconnection layer 
is defined in \cite{zhang_value_2014} as part of the Internet content delivery activity 
where the marketed product is the transportation of content 
over the Internet. Hence, cash flow does not depend on the content but only
on the bits transferred.} for two content delivery models (CDN and ICN), 
analyze the impact of ubiquitous encryption on such models (in terms of cash flow and control/performance)
and outline possible future architectures/mechanisms that could be advantageous for the network stakeholders.\\

In section~\ref{sec:cdn} and section~\ref{sec:full-collaboration} we present the current CDN model and study its dynamics in 
the absence and presence of encrypted traffic respectively. Following the thrust of recent research papers, we then introduce
the ICN model (section~\ref{sec:icn}) and propose an additional implementation (section~\ref{sec:proposal}) which is promised
to challenge the CDN model. We t
In section~\ref{sec:performance} we compare the presented CDN and ICN models in terms of a few key parameters/business-enablers
and try to figure out which model will eventually stand out in the long run. Finally, section~\ref{sec:performance} will serve to 
recap the main
findings of this paper.

%
%
%
%

\section{CDN model background}
\label{sec:cdn}
\begin{figure}[t]
\begin{minipage}[l]{0.45\linewidth}
\hspace{-1cm}
\includegraphics[width=1.1\linewidth]{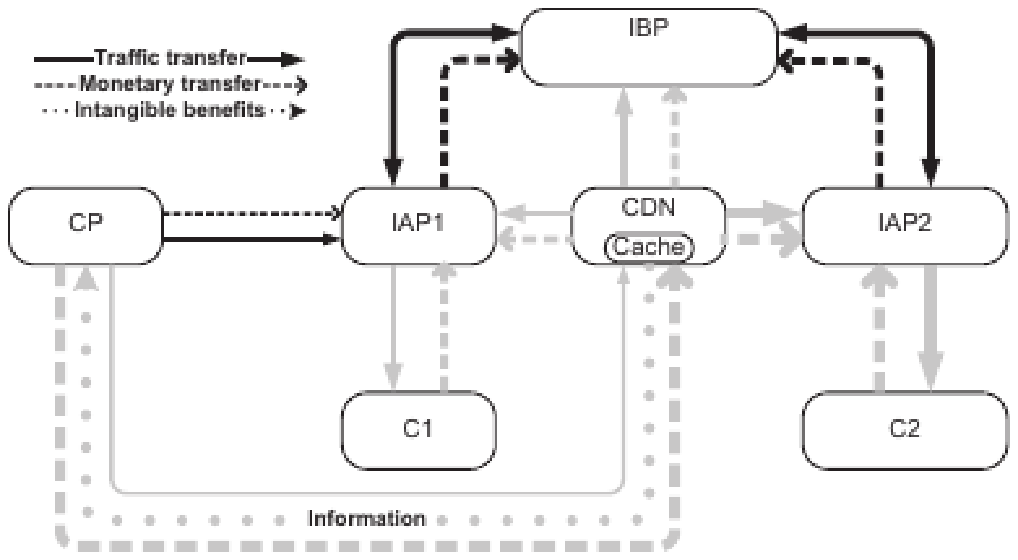}
\end{minipage} \hfill
\begin{minipage}[r]{0.45\linewidth}
\hspace{-0.6cm}
\includegraphics[width=1.1\linewidth]{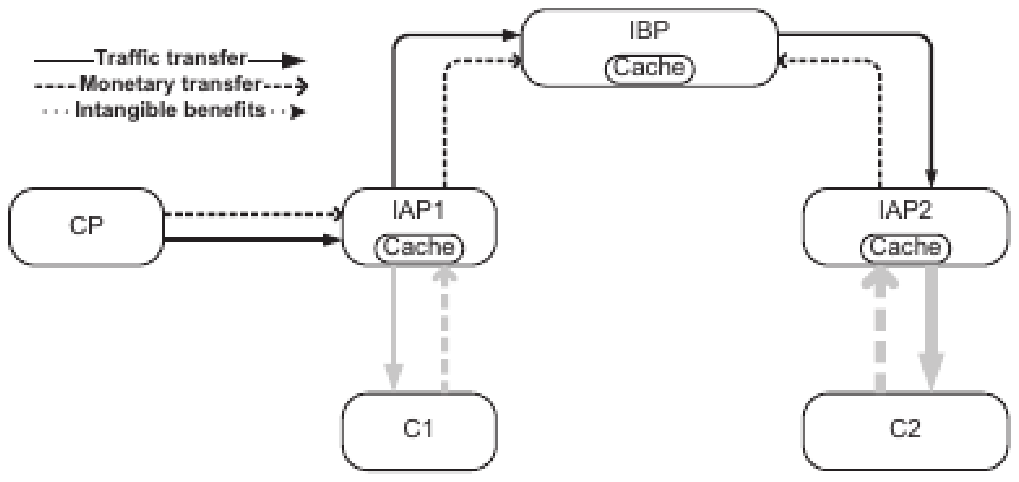}
\end{minipage} \hfill
\caption{Value networks of CDN model (left) and ICN model (right). Figures are drawn from \cite{zhang_value_2014} and depict 
the interactions between Content Provider (CP), Internet Access Providers (IAP1 and IAP2), Content Delivery Network (CDN),
Internet Backbone Provider (IBP) and end-users (C1 and C2). IAP1 (aka "content" heavy IAP) provides hosting and network access for 
end-customers and commercial companies that offer content (such as Google and Yahoo). Conversely, IAP2 (aka "eyeball" heavy IAP)
supports the last-mile connectivity. To notice the reduced
circulation of money with the ICN model wrt the CDN model.}
\label{fig:cdnicn}
\end{figure}

In the CDN model (Fig.~\ref{fig:cdnicn}-left) the CP outsources (at least part of\footnote{Typically, the delivery 
of heavy, popular and delay-sensitive content, such as video traffic, is served by CDNs.}) the content delivery to a CDN provider, 
which figures as a new stakeholder in the original client-server model. The CP can be connected to the CDN either through 
the Internet Access Provider (IAP) the CP is attached to (named IAP1), or directly through a peering
link as is shown in the figure. In the CDN model, some of the content requests are served from the CDN's servers. 
As a result, the traffic volume
between the CP and IAP1 decreases. Nevertheless, this does not necessarily reduce the CP's costs for hosting and serving the content, 
as it pays the CDN for the
volume of traffic delivered to end-users. Thus, the monetary transfer from the CP to the CDN is proportional to the traffic transfer
between the CDN and IAPs rather than the traffic transfer between the CP and the CDN. For example, \cite{drpeering}
compared four video distribution models (Internet transit, CDN, hybrid of transit and peering, and P2P) with prices from
2006 and concluded that the CDN model is the most costly for the CPs, especially with low traffic 
volumes\footnote{Typical CDN prices were ranging from 2 USD per Mbps to 11 USD per Mbps in 2012 \cite{_cdn_????}}. \\

If thinking in terms of large traffic volumes instead, the CDN may charge less for the traffic than an IAP. For example, the
transit price for 2012 was 2.34 USD per Mbps \cite{drpeering} and a content provider with 400 Mbps a month
was paying as low as 2 USD per Mbps \cite{_cdn_????} to the CDN provider\footnote{Note that the CP still needs to get the content to the 
CDN - usually over the Internet. So the CP has that cost too.}. \\

A CP's incentive to use CDN stems from
the unique benefits that the CDN provides: namely, better scalability, decreased load on CP's origin server and improved end-user 
experience, especially to customer 2 (C2 in the figure). The CDN is also able to collect valuable usage information, which 
is shown as an intangible benefit in the figure. The CDN may then resell these analytics to the CP,
thus thickening the cash flow arrow from the CP to the CDN in the figure. Moreover, if a CDN partnered with a Mobile
Network Operator (MNO), they 
could give richer analytics information including anonymized customer data.\\

In the emergence of the CDN model, IAP1 loses part of its revenues received from the CP, but also saves in the transit
costs (as does IAP2) since the traffic is delivered to consumers over the peering links between the CDN and the IAPs.
Traditionally, CDNs and IAPs had settlement-free peering agreements \cite{drpeering} and IAPs allowed CDNs to collocate
servers in the IAPs networks for free. However, due to the increase of heavy traffic going through the CDN (typically
video contents), IAPs increasingly
charge CDNs for peering \cite{clark_interconnection_2011}. In Fig.~\ref{fig:cdnicn}-left, the fee charged by the 
IAP from the CDN corresponds to the traffic
volume going through the CDN and the IAPs. However, in reality, the terms of paid peering agreements vary and may not
depend strictly on the traffic volume. Other terms can be used, e.g., if the traffic is encrypted. This scenario is discussed
in section~\ref{sec:full-collaboration}.

Finally, from the Internet Backbone Provider (IBP) perspective, 
the CDN can generate additional traffic by stimulating content consumption from subscribers; however the CDN may 
compete for the IBP's content delivery market and decrease the IBP's revenue. The negative impact of the CDN
model depends on the number of peering agreements between the CDN and the IAPs. If the coverage of the CDN's peering is
extensive, less traffic and money is transferred between IBP and IAP1/IAP2/CDN.

\section{QoE optimization: a matter of control}
\label{sec:full-collaboration}

The wider use of encryption is leading to a greater use of HTTPS between web browsers (i.e., end users) and content servers,
thus forcing the IAP to play the role of a dumb pipe in the worst case scenario. Since the amount of encrypted traffic over the Internet is 
designated to increase further in the years 
to come, most IAPs are starting
to take action so as to fit the Internet of encrypted things. In this new context, an IAP's ultimate goal is to keep 
an acceptable level of traffic control over its network so as to be able to keep guaranteeing a certain 
protection and QoE to their 
customers\footnote{Customers protection is here intended in terms of content filtering/monitoring. This is in fact an important operation
if, e.g., the customer profile is set to "child".}. To this 
end, we single out several possible business development directions and outline advantages and drawbacks of each of them.

\subsection{Legitimizing eyeball IAPs' HTTPS traffic inspection}
This is possible in two different ways:
by terminating HTTPS traffic via a man-in-the-middle like approach (which can be performed by the operator itself or by means 
of a dedicated trusted proxy server).
With such mechanism, Mobile Network Operators (which are eyeball IAPs) would be able to fully control and optimize their network. 
A few promising ideas in this direction (mainly related to caching operations)
are reported in subsection~\ref{subsec:appcache}. Note that terminating HTTPS traffic via man-in-the-middle approach,
although volatile\footnote{Techniques such as Certificate Transparency, DANE and operating 
system control of certificate stores can detect and prevent man-in-the-middle certificate spoofing.} and impractical for large volume of 
traffic (as underlined in the Introduction), is simple to put in place (e.g., via Vodafone 
CleanPipe or a Cisco proxy server) and can be of help when IAPs and operators are given the governmental mandate to inspect encrypted traffic 
in order to support national security as well as the fight against international terrorism (like Islamic State).

\subsection{CDN-IAP full collaboration}
\label{subsec:full-collaboration}

In the case of encrypted traffic, information available in the access network such as 
caching capabilities, instantaneous throughput and users' mobility can be put at stake and sold to the CDN or exchanged with 
other information . This business
model is known as Network as a 
Service (NaaS). The benefits of 
a full collaboration between mobile operators (which are eyeball IAPs) and CDNs would therefore include the renting of operators' servers resources 
for CDN Edge Servers, the exchange of non user-specific information such as average cell throughput\footnote{Mobile throughput guidance 
mechanism \cite{sprecher_mobile_????} has brought good network load and latency performance gains in prototype lab trials.}, as well as
TCP level optimization. In addition, such an approach would also be suitable for communications over QUIC (Quick  UDP  
Internet  Connections)\footnote{Despite Google has already deployed QUIC
protocol in their servers and has a client implementation in
the Chrome web browsers (note: QUIC has been developed by Google), it is still an experimental protocol.}, since it would provide valuable information to the QUIC server for optimizing its statistical techniques,
namely packet pacing (with ongoing bandwidth estimation) and proactive speculative retransmission (sending duplicate copies of 
the most important packets).\\

On the flip side, we would like to stress the fact that in the perspective of a NaaS approach, IAPs
should carefully evaluate any potential margin squeeze to avoid conflicts with the CDN. In fact, a higher price charged to the CDNs would have implications for CPs and, 
in some cases, for the final users. For instance, Netflix could be forced to increase its monthly fee, this causing less users to
buy its services, this causing less traffic through the IAP and the CDN, this causing even less money for the CDN.\\

\subsection{TCP level optimization}
A possible idea to let the IAP reduce the traffic flowing across its network 
is to implement a TCP freeze mechanism in the radio access nodes (such as LTE evolved Node BSs or HSPA+ BSs) that follows the status
(radio channel quality, handover, etc.) of the user plane connections, cell load or any other relevant measurement to freeze the selected
TCP connections with Zero Window Advertisements (ZWA) messages \cite{deak_tcp_2013}. 
Thus, TCP  senders are "frozen" whenever transmission problems are detected or
predicted and
the
operation
of the  TCP are resumed when  the  system  is  able to operate efficiently again. A TCP sender is frozen if it
receives an ACK segment with ZWA. The latter forces  the TCP  sender  to  suspend all retransmission
timers   and  enter persist  mode,  which  lasts until  the source receives  a Non-Zero
Window Advertisement (NZWA)  segment,  i.e., a TCP  ACK with  positive advertised window  size.  In between, the  TCP
sender does not send any new data and the Retransmission Timeout (RTO) timer  will not
expire even if the persist mode lasts much longer than the RTO.
When  the  NZWA is received, the data transmission continues with  the  same  cwnd  size  (i.e.,  at  the  same  sending  rate)  as
before  the freeze.  Therefore,  a timely  freeze  will minimize the probability
of  a TCP  slow  start  caused  by  poor  radio  channel conditions and  will  allow  the  TCP  to
continue with  the  same speed  after the coverage
problem is over instead  of recovering through  slow  start.\\

Note that such solution does not require the use of 
dedicated messages (such as ICMP) or other commands in order to freeze the TCP senders. Moreover, the proposed functionality is
transparent to the TCP layer and no additional mechanism is needed at the TCP sender or receiver sides. It is shown in \cite{deak_tcp_2013}
that the TCP freeze-based mechanism brings striking gains in terms of average outage time (20-30\% performance gain) and throughput
(15\% lower download time) in the case of coverage holes.\\

\subsection{Content-heavy IAPs becoming CDN-0}
\begin{figure}
\begin{center}
\includegraphics[width=0.7\linewidth]{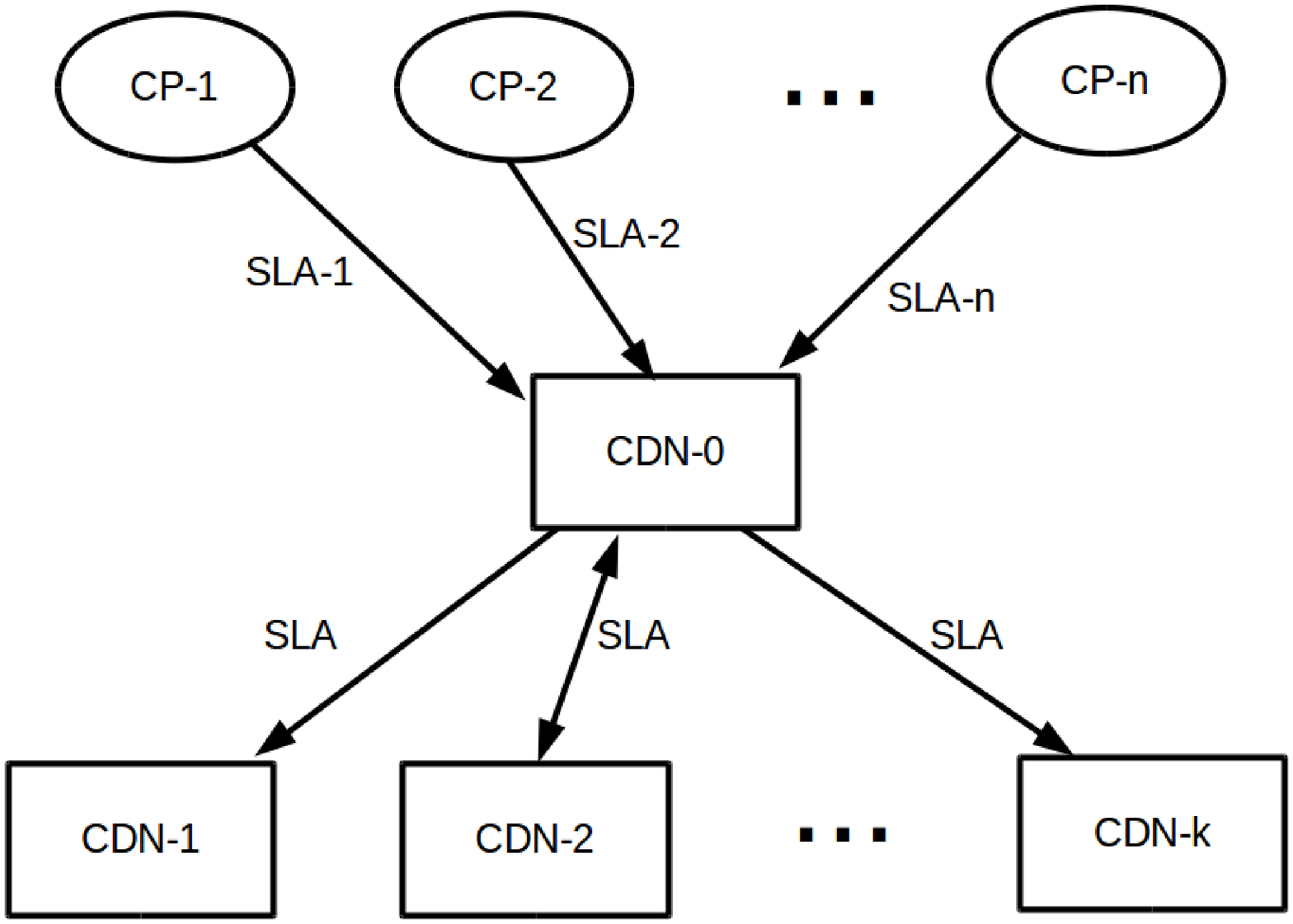}
\end{center}
\caption{Multiple content providers sign business contracts with one authoritative CDN with various 
Service Level Agreements (SLA); 
the authoritative CDN-0, as a customer, rents
services from other CDNs, besides using its own infrastructure.}
\label{fig:cdn0}
\end{figure}

In \cite{wang_content_2014}, the authors advise that one could leverage content multihoming to decrease prices for 
low traffic CPs.
This is possible by letting each content provider sign only one contract with an authoritative CDN, say CDN-0, and by letting CDN-0 
impair
the potential performance trap of individual CDN. Thus, 
CDN-0 acts as a proxy, aggregating demands from registered CPs and acquiring
the maximum discounts from other CDNs as a “giant CP”.
Such scenario is depicted in Fig.\ref{fig:cdn0}.\\

To fulfill its business commitment, CDN-0 could either use its own 
infrastructure or rent services from other CDNs, which can be indispensable for certain data transfers and certain 
types of content. This holds true 
regardless of the content
being encrypted or not, though there might be regulatory hassle if CDN-0 is an 
“eyeball” heavy IAPs (IAP2 in Fig.~\ref{fig:cdnicn}-left). 
In the latter case, the operation could be justified
by bringing Internet democratization arguments, i.e., by arguing that small CPs unfairly have to pay higher prices for CDN
services while getting less performance guarantees wrt large CPs which sometimes also own a CDN. \\

\subsection{Large eyeball IAPs becoming a CDN}
\label{subsec:IAPCDN}
It has been formally shown in \cite{agyapong_economic_2013}
that the only viable architecture maximizing social welfare (in terms of profitability of the stakeholders) and 
large eyeball IAPs\footnote{Here for large eyeball IAPs 
we have in mind eyeball IAPs with multi-country coverage (and thus a large amount of potential customers). This can 
be a single large IAP or a federation of IAPs.} revenues is the one where
large eyeball networks provide the caching infrastructure through a federation
(thus 
becoming de-facto a CDNi), and a third-party entity serves as a transaction broker between the CP and various eyeball
IAP networks.
Note that this model is proven to be the only way 
for large eyeball networks to appropriate most
of the value created from caching. Even, it would not be a deal-breaker for classical CDNs (like Akamai), since   
medium and small eyeball networks would still find hosting classical CDNs beneficial. 
We will then show in subsection~\ref{sec:proposal} how this 
can further evolve 
by leveraging the positive effect of a new technology such as ICN.\\ 



\subsection{Application level caching operations}
\label{subsec:appcache}
\begin{figure}
\begin{center}
\includegraphics[width=0.9\linewidth]{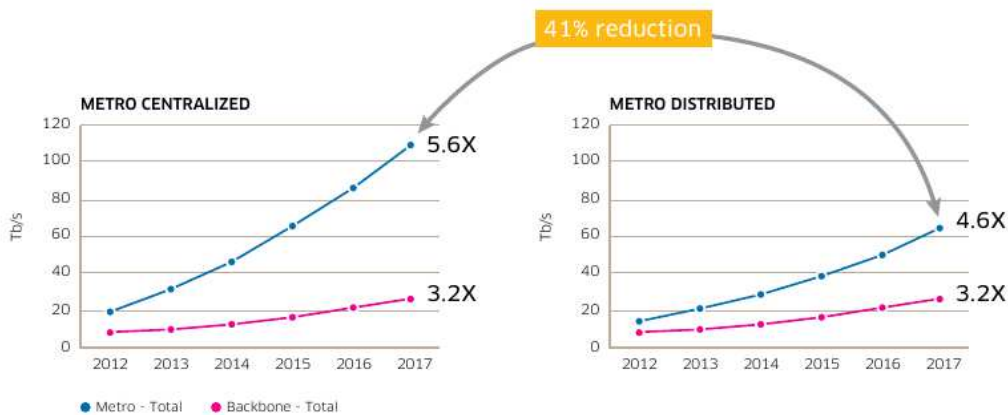}
\end{center}
\caption{Metro distributed caching provides bandwidth savings \cite{bell_labs}}
\label{fig:metro_traffic}
\end{figure}

From what has been said, one can easily infer that position, control and size of cache are among the assets that can be 
more easily monetized, especially in CDN models.
Hence, it is important for business people 
to keep an eye on the technical findings on caching management and operations. With this in mind, we report 
here below some of the most promising
research directions related to the matter:
\begin{itemize}
 \item {\bf Cache-enabled base stations}: In a recent paper \cite{bastug_cache-enabled_2014}, it is formally demonstrated that
significant gains in terms of outage probability and average delivery rate are possible by having
cache-enabled {\it small base stations}. Hence, telecom operators can either deploy more base stations or
increase the storage size of existing deployment in order to achieve a certain QoE level. Similar recommendations can be
derived from \cite{imbrenda_analyzing_2014} for LTE-BSs and serving gateways (which serve a number of base stations).
The benefits of such a distributed caching architecture is then further investigated in \cite{bell_labs} and the main 
result is reported in Fig.~\ref{fig:metro_traffic}. Note however that caching one-hop back from the SBSs may prove financially
and operators will likely have to find a trade-off between performance and costs.  
\item {\bf Exploit mobility in caching}: Caching operations from Radio Access Provider's edge can be enhanced by 
exploiting users' mobility.
In a 5G-like scenario for instance, given predictions about mobility and request patterns of users, it is possible to 
determine the content to be usefully cached
at each small base station (SBS) to minimize macro-cell's load. Thus, SBSs that are sequentially visited by users should cooperate 
when caching contents. For example, a user who is moving across two small cells and is in the process of downloading a file, 
should be able to download
half of the file from each SBS (if the file parts are appropriately cached).
\item {\bf Combine multicast with caching}: The idea is to optimize the delivery of content to multiple users requesting the same 
content in close times.
This mostly applies to
scenarios of large number of people being in one area and sharing a common interest, e.g, sport game in a 
stadium where people want to know live statistics and scores from other games. In a 5G-like scenario for instance,
users' requests
for the same file are collected and served periodically by a single multicast transmission from the macro-cell base station (MBS). In
order to avoid such (costly) MBS transmission, all the SBSs that receive request for the same file should have it cached. 
Hence, given predictions about user request patterns and period of multicast service,
the challenge is to determine the content to be cached at each SBS to minimize macro-cell's load. 
\item {\bf Predictive cache}: Contents can be downloaded before the users actually require them. The gain of such an approach
has been demonstrated on real data provided by BBC iplayer set-top-box (paper under review). 

\end{itemize}

\section{ICN model background}
\label{sec:icn}
Though several value networks can be constructed for ICN,
we concentrate on a value network,
where IAP transparently caches named content similarly to web caching. As shown in Fig.~\ref{fig:cdnicn}-right, caches are then
located at and controlled by the ISPs. In this scenario, the CP loses the overall control on the content
delivery as the ICN system transparently caches the content and serves the content requests from the optimal, nearby
locations. As a result, traffic and its related cash flow between the CP and the IAP1 decrease to the same level as in the CDN
models.
Due to the optimal cache selection, also the traffic volume through the Internet backbone decreases. Therefore, the monetary transfer between the IAPs and the IBP reduces, which may lead to resistance from the
IBPs and lower their interest to participate in the ICN deployment. The increased costs for the IAP from building and
maintaining the cache infrastructure are not illustrated in the value network, because only traffic related costs are modeled.
However, these costs are taken into consideration in Section \ref{sec:performance}.\\

\subsection{Challenges of Information Centric Networking}
\begin{figure}
\begin{center}
\includegraphics[width=\linewidth]{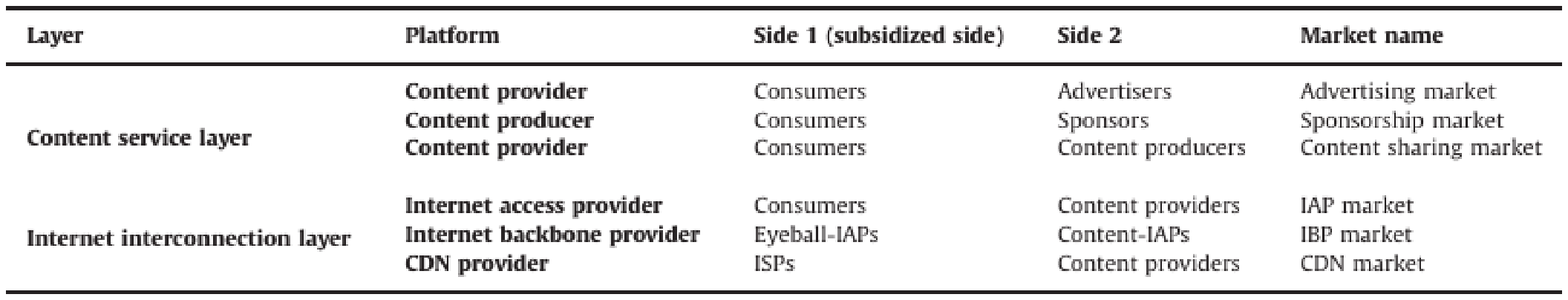}
\end{center}
\caption{A two-sided market is defined as a market with two distinct sides that are interlinked (through a platform) and where not only the 
overall price level matters, but also the price structure between the two sides \cite{zhang_value_2014}. The demand asymmetries
between interlinked markets typically lead to skewed pricing with one side charged more than the other, which sometimes can be priced
even under production cost. A typical example is the free software market, where the software is distributed free to consumers (because
they are more price-sensitive) and advertisers are priced depending on the popularity of the software/content. In this case, consumers
are said to be subsidized.
In the IAP market case instead, consumers are the subsidized side and CPs are the other side.}
\label{fig:two-sided}
\end{figure}

In this subsection, we identify and discuss several challenges for the deployment of ICN. 
First is the two-sided pricing in the CDN market (CDN model two-sided market and pricing is detailed in Fig.~\ref{fig:two-sided}), which causes
CDN providers to consolidate and become Internet-wide CDNs\footnote{Such consolidation has taken the form of
a standardization work (named CDNi) at IETF \cite{brandenburg_models_????}. However, 
such  standardization  activity  struggles  with  attracting the main CDN players, 
so raising concerns about the success of the integration process.}. A network wide CDN or interconnected CDN providers (i.e. via CDNi) would have the same
coverage as the ICN model and, thus, can be a direct competitor to ICN. In addition, a CDN provider may move towards ICN
by placing more cache servers into the network and closer to the end-users, especially if a tight collaboration with the ISP 
is in place (see subsection~\ref{subsec:full-collaboration}).\\

The CDN model, being a two-sided market, has a central platform that coordinates the service provisioning and offers
service level guarantees. In addition, the CDN provider offers usage statistics to the CPs as a service. The proposed value
network model for ICN does not have a unique two-sided market with a central platform that can offer guaranteed service
levels and value added services. As CPs value both service level guarantees and usage statistics, the ICN model needs to solve
these issues to gain acceptance from CPs. In addition, for the advertisers, who are a revenue source for CPs, being able to
identify each individual user is important, but caching makes the identification less straightforward. To solve this problem,
IAPs could collect usage and user information for the CPs. However, ICN's ability and IAP's willingness to provide such
services is uncertain.\\

The CP's control over its content is another issue in ICN, especially with dynamic contents such as targeted marketing.
This means that conflicts may arise between CPs and IAPs: IAPs may not have the incentive to update content caches very
frequently while CPs may wish the content to be up to date. Legal measures, business agreements and monetary incentives
may help in solving this conflict and the other limitations of the ICN model identified earlier. Nevertheless, these limitations
have to be solved before ICN can prosper.
The central platform also makes charging of the service simpler. In the ICN model, it is unclear who can monetize the
service offering and who should be charged for the service, and this could lead in the worst case scenario
to accusations of "loss of access contro" of copyright content by the CP, especially where content is ad-funded.
The CDN market's analysis suggests that CPs, compared to the
ISPs, have higher willingness to pay for better-than-best-effort content delivery. Despite this, ISPs'
willingness to deploy ICN is more crucial as they control the network locations, where cache servers should be located. 
One potential platform
could, thus, be the ISPs, if they can pass the extra costs to the consumers or CPs. For example, the name resolution system
(i.e., rendezvous) in NetInf and publisher/subscriber can be considered as a two-sided market, which matches the requesters to the
publishers. A natural owner of this function is the ISP. However, name resolution
remains difficult to
monetize for an ISP. An idea could be to include it as part of an SLA that 
matches the customer to better-than-best-effort content hosts via ICN.\\

IAPs' willingness to invest in ICN can be determined by summing up the net benefit of reduced off-net 
traffic\footnote{Traffic is said to be on-net, if both end-points are located in the same IAP's network; and off-net, if
the end-points are located in different IAPs networks. As a consequence, on-net traffic is cheaper for an IAP to deliver. 
Off-net traffic can be delivered either through peering or transit agreements, where peering typically is the cheapest option for
an IAP.}, extra
investments and other costs (e.g., commercial reshaping might be needed since IAPs may not currently have the human
resources or organization to deal directly with CPs). However, the ICN model considered in this section may not be able to compete with the CDN
model, in which the IAPs avoid additional investments but still enjoy the same transit cost savings accompanied by possible
payments from CDN providers in case of paid peering. Thus, the IAPs may need further incentives to invest in ICN from
possible additional revenues and monetization schemes which would counterbalance the missed revenues wrt the CDN model (e.g., the ones 
from charging CDNs to peer). For example,
an eyeball-IAP
has the incentive to invest in caching, provided that it can charge high enough prices from the CPs. On the other hand, ICN may enable
a shift towards more content-oriented interconnection agreements, where not only the transferred data volume but also the
availability of the content from the caches of a peering partner is taken into consideration. If the value of
the available content is higher than its transfer costs, this new way of peering may serve as additional incentive for the IAPs
to invest in ICN.\\

Finally, based on the value network analysis, an IBP's willingness to add cache servers to its network is very low, but it may
be stimulated by the significantly falling Internet transit prices. Since transit revenue has traditionally been the major income 
for IBPs, IBPs are now shifting to other revenue sources such as providing CDN services. Thus, IBPs may also be open to the 
possibility of finding viable revenue models from ICN.

\section{ICN federation model}
\label{sec:proposal}
After having gone through the precedent section, one can infer that the considered cache 
is at the same time an advantage (from a technical
perspective) and a disadvantage (from a business size perspective). In fact, the self-organizing transparent cache featured in most ICN 
implementations is difficult
to monetize as it is a common asset that works like a black box (meaning that none can control it). \\

\begin{figure}
\begin{center}
\includegraphics[width=0.7\linewidth]{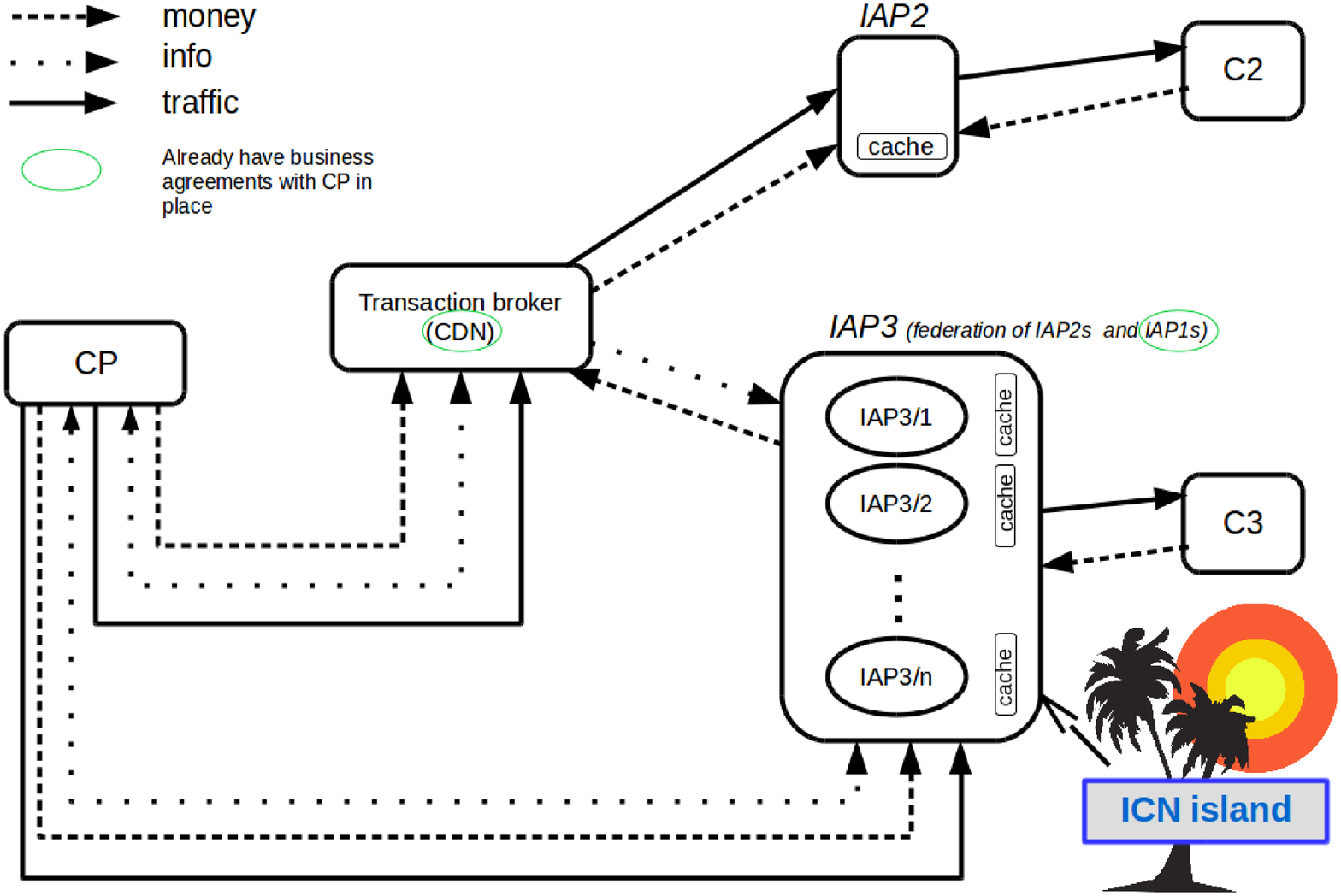}
\end{center}
\caption{Value network of the ICN federation model.}
\label{fig:IAP3}
\end{figure}

In this section, we draw inspiration on \cite{agyapong_economic_2013} and propose a possible business model based 
on ICN with commercial cache, meaning that the content allocation across 
caches is decided by a trusted authority, which we refer to as transaction broker. The model is similar to the one outlined 
in subsection~\ref{subsec:IAPCDN}, except for: 
\begin{itemize}
 \item The CDN (e.g., Akamai) is the transaction broker. This new trusted role brings additional revenues to the 
 CDN (from the ICN federation) and could thus facilitate the transition towards such model.
 \item  The ICN federation members deploy and manage their 
own ICN networks
separately, but still enjoy the benefits of a common entity that harmoniously encompasses CPs, end-users and CDNs. 
 \item The federation is ICN-oriented. Therefore, this should be considered as a long term plan for the IAPs. Thereby,
 this can be coupled with short/medium term strategy adjustments which can include the actions outlined in section~\ref{sec:full-collaboration}.
\end{itemize}
The value network of the ICN federation model is depicted in Fig.~\ref{fig:IAP3}. The CP has here potential control over distribution
and location of its contents (in the figure, through the information flow going from the CP towards IAP3 and transaction broker), 
meaning that the CP can bias 
the allocation procedure chaired by the transaction broker.
Note that for ease of read, the money and traffic flow
between the CDN and the IAP3 (i.e., the ICN federation) is not reported (though is definitely present), nor is the 
IBP. The latter could be possibly included into the ICN federation, though a role as standalone connector is also possible 
(the IBP will have to carefully evaluate new margins and return on investment).

\section{Performance evaluation}
\label{sec:performance}
One could argue: Is it sustainable for operators to enrich all their BSs with caching capabilities? 
Can their revenues scale with the dramatic predicted traffic increase which will come with 5G? Can in-house network optimization
counterbalance the increasing operational costs, especially with encrypted traffic? Or is it better to share such costs with the CDN by letting the latter take part 
in the access network management operations? How much is worth the full control over the IAP network? Can ICN be an answer 
to all these questions?

\begin{figure}
\begin{center}
\includegraphics[width=\linewidth]{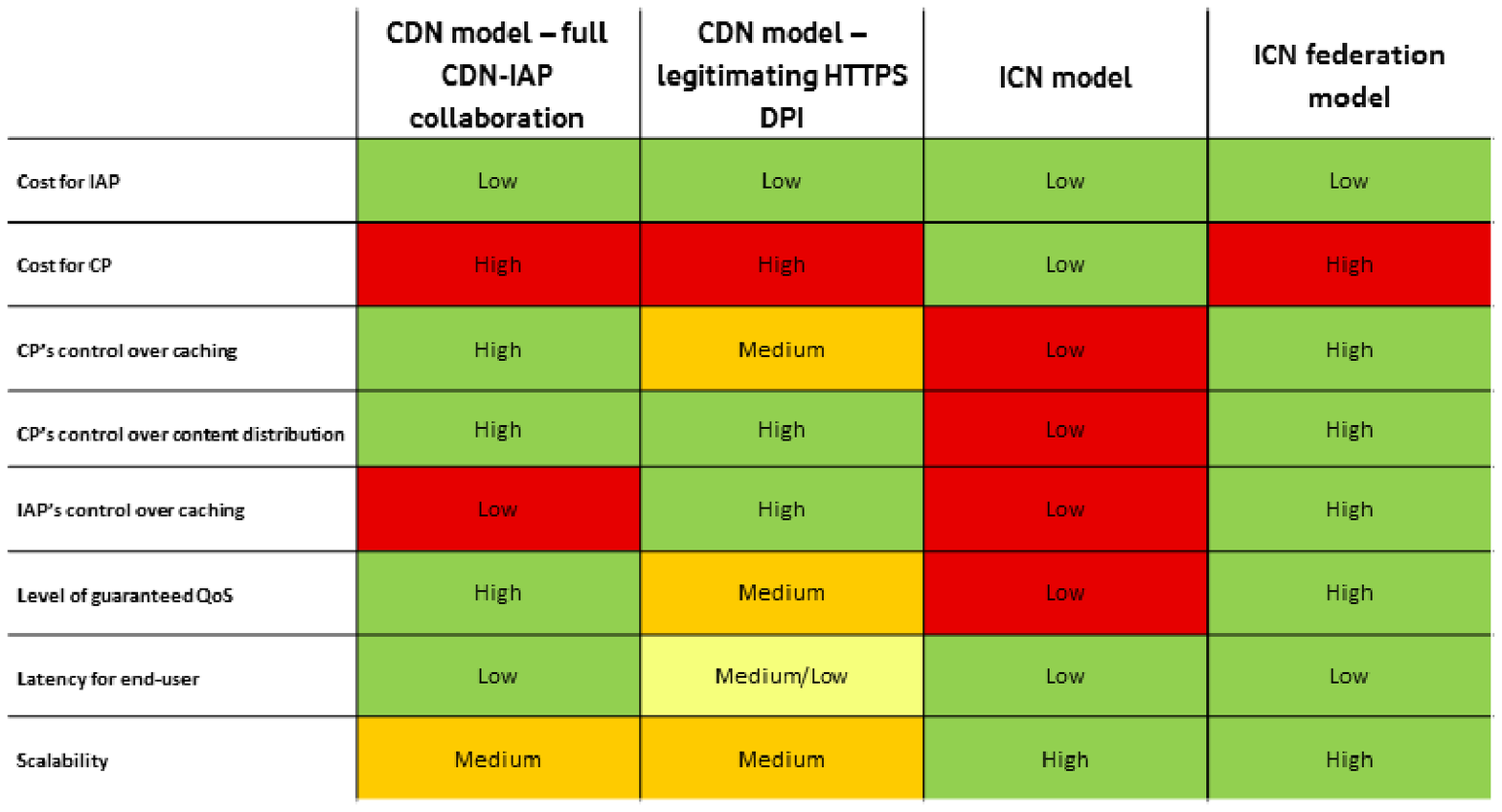}
\end{center}
\caption{Performance comparison between CDN and ICN in terms of secure traffic (i.e., we consider encrypted traffic for the CDN
case).}
\label{fig:CDN_ICN_HTTPS}
\end{figure}

In an attempt to give an answer to these questions, we have compared the performance of CDN and ICN models in terms of a few key
parameters, namely latency for end-user, control (on caching and content distribution), network scalability
and potential regulatory hassles. 
The results are summarized in Fig.~\ref{fig:CDN_ICN_HTTPS} and some comments are provided below. \\

CP's control over caching (i.e. where to cache, what to cache and how often content is updated in
caches) is higher in the fully collaborative CDN model and ICN federation model 
due to the business agreements in place between CPs and CDNs/ICN federation.
It is on the contrary low in the 
classical ICN model and medium in the deep packet inspection (DPI) case, because the CPs and ISPs do not have
caching related business agreements.\\

Similarly, the CDN model and ICN federation model best
implement guaranteed service quality to the end-user\footnote{We suppose that deep packet inspection
could result in delays that would impact QoS and scalability.
In the meantime, we assume medium scalability and level of guaranteed QoS.}). 
Since providing a certain level of service quality to end users is important for a CP, 
the cost savings ensured by other models may not compensate for the possibly
lower service quality and control. In addition, the QoS level is low in the ICN model, because of uncontrolled transparent
caching, though different ICN implementations such as the proposed ICN federation model may offer different 
levels of QoS \cite{agyapong_economic_2013}. \\

From the IAPs perspective, the most important factors are the QoS for the end user and the control 
over the encrypting contents flowing across the IAP network. In this regard, the DPI approach and ICN federation 
are the most suitable solutions.\\

Finally, looking at the network as a whole, one of the critical factors is scalability. In this respect, 
it is safe to assume that ICN models scale better with the traffic, due to the distributed
nature of the network, which is preferable for all the stakeholders.\\

%
%

\section{Conclusion}
\label{sec:conclusion}
Content Delivery Networks (CDN) have emerged in recent years as the reference solution for secure and fast
content delivery over the Internet. Such model currently prospers because it's been able to reduce the coordination
problem of end-to-end quality of service (QoS) by leveraging the demand asymmetry between Internet access providers
(IAPs) and CPs. Thus, the successful deployment of two-sided pricing has made CDNs attractive to both IAPs and CPs and has caused
the CDN providers to consolidate and become stronger.\\

With this in mind, we have carried on an analysis of the content delivery market by focusing on a few key factors, namely ubiquitous encryption, traffic
increase, network scalability, latency needs and network control. We have shown that 
if one or more of such key factors approach a critical level the current CDN model might be destabilized and not 
survive in the long run. \\

Despite that, we expect a consolidation of the present CDN model in the short term, with the flourishing of the peering business and the 
consequent emergency for tighter collaborations
between IAPs and CDNs. Differently, in the long term we see the possibility that large IAPs will be challenged by the lack of control over 
encrypted contents and by 
the huge amount of traffic coming from  
bandwidth-hungry new-generation devices and video-oriented services. \\


The Information Centric Network (ICN) model, considered by some as an evolution of the CDN model, is based on a new networking
paradigm characterized by unique
content 
naming and optimized caching features. Although this may be seen as a prominent concept
for improving the scalability of the Internet, its design is facing 
technical and, especially, business challenges. We have therefore taken up these challenges and proposed an implementation of 
ICN - characterized by the presence of commercial cache and transaction broker - which is promised to maximize large IAPs
revenues and to guarantee an equilibrium among the network stakeholders. \\

The results presented in this paper 
can be readily utilized by large IAPs to arrange a timely roadmap aimed at
anticipating scalability and control 
problems in the access and core network. Such problems can be reconducted to a performance/cost
trade-off and can be solved by means of cache-oriented solutions and, eventually,
large-scale 
ICN deployments. However, the standardized global deployment of ICN requires that the incentives of stakeholders are aligned. 
To this end, the CPs (i.e., the
decision drivers) should be willing to pay for lower delay in content delivery. \\

\newpage\cleardoublepage

\bibliographystyle{abbrv}  
\bibliography{encrypted} 

\begin{thebibliography}{10}
\bibitem{_most_????}
{Most {Internet} traffic will be encrypted by year end. {Here}'s why. -
  {Fortune}}.
\end{thebibliography}

\end{document}